\documentclass[vecphys]{svmult}		% vectors in bold face
\usepackage{amsmath}
\usepackage{makeidx}         % allows index generation
\usepackage{graphicx}        % standard LaTeX graphics tool
\usepackage{epsfig}          % another way to include figures
\usepackage{multicol}        % used for the two-column index
\usepackage[bottom]{footmisc}% places footnotes at page bottom

\makeindex             % used for the subject index
\begin{document}
%%%%%%%%%%%%%%%%%%%%%%%%%%%%%%%%%%%%%%%%%%%%%%%%%%%%%%%%%%%%%%%%%%%%%

\newcommand{\beq}{\begin{equation}}
\newcommand{\eeq}{\end{equation}} 
\newcommand{\beqa}{\begin{eqnarray}}
\newcommand{\eeqa}{\end{eqnarray}}
\newcommand{\ba}{\begin{array}}
\newcommand{\ea}{\end{array}}

\title*{Contact intensity and extended hydrodynamics 
in the BCS-BEC crossover} 
\titlerunning{Contact intensity and extended hydrodynamics}
\author{Luca Salasnich} 
\institute{Dipartimento di Fisica e Astronomia 
``Galileo Galilei'' and CNISM, \\
Universit\`a di Padova, Via Marzolo 8, 35122 Padova, Italy \\
\texttt{luca.salasnich@unipd.it}} 

%\pacs{03.75.-b; 05.30.Fk; 71.10.Ca}

\maketitle

\abstract
{In the first part of this chapter we analyze 
the contact intensity $C$, which has been introduced 
by Tan [Ann. Phys. {\bf 323}, 2952 (2008)] and  
appears in several physical observables 
of the strongly correlated two-component Fermi gas. 
We calculate the contact $C$ in the full BCS-BEC 
crossover for a uniform superfluid Fermi gas 
by using an efficient parametetrization of the ground-state energy. 
In the case of harmonic confinement, within the Thomas-Fermi approximation, 
we derive analytical formulas of $C$ in 
the three relevant limits of the crossover. 
In the second part of this chapter 
we discuss the extended superfluid hydrodynamics 
we have recently proposed 
to describe static and dynamical collective properties 
of the Fermi gas in the BCS-BEC crossover. In particular 
we show the relation with the effective theory for the Goldstone field 
derived by Son and Wingate [Ann. Phys. {\bf 321}, 197 (2006)]  
on the basis of conformal invariance. By using our equations 
of extended hydrodynamics we determine nonlinear sound waves, 
static response function and structure factor of a generic superfluid 
at zero temperature.} 

\section{Contact intensity} 

It has been proved by Tan \cite{tan1} that 
the momentum distribution $\rho_{\sigma}(k)$ in an arbitrary system consisting 
of fermions in two spin states ($\sigma=\uparrow,\downarrow$)
with a large scattering length has a tail that falls off as 
\beq 
\rho_{\sigma}(k)\sim {C\over k^4} 
\label{define}
\eeq
for $k\to \infty$, where $C$ is the so-called contact 
intensity \cite{tan1}. 
Here large s-wave scattering length $a$ means that $|a|\gg r_0$, 
where $r_0$ is the effective interaction radius. Under this condition  
Tan \cite{tan1} has shown that $C$ is related to 
the total energy $E$ of the Fermi system by the rigorous expression 
\beq 
C = {4\pi m a^2\over \hbar^2} {dE\over da} \; ,  
\label{figa}
\eeq
where the derivative is taken under constant entropy and, 
in general, $C$ depends on the number $N$ of fermions, 
the scattering length $a$ and the parameters of the 
trapping potential \cite{tan2,tan3}. 
Remarkably, Eqs. (\ref{define}) and (\ref{figa}) work also at finite 
temperature and in this case $C$ will be a function 
of $T$ \cite{tan2,leggett}. 
Tan has also derived, for finite scattering lengths, 
a generalized virial theorem and a generalized pressure relation 
where the contact $C$ appears \cite{tan3}. 
The contact intensity $C$ appears also in other physical observables 
of the strongly correlated Fermi system. For instance, the radio-frequency 
spectroscopy shift is proportional to $C$ 
\cite{zwerger,zwierlein,pieri,pieri2}, 
and the same happens to the photoassociation rate \cite{castin}. 
Very recently, it has been shown that the contact $C$ gives 
the asymptotic tail behavior of the shear viscosity 
as a function of the frequency \cite{braaten3}. 

Using the methods of quantum field theory, Braaten and Platter 
have rederived \cite{braaten1,braaten11} the Tan's universal 
relations \cite{tan1,tan2,tan3}. 
In addition, they have shown that the contact intensity can be written as 
\beq 
C = \int g^2 \langle {\hat \psi}_{\uparrow}^+({\bf r}) 
{\hat \psi}_{\downarrow}^+({\bf r}) 
{\hat \psi}_{\downarrow}({\bf r}) {\hat \psi}_{\uparrow}({\bf r}) 
\rangle \ d^3{\bf r} \; , 
\eeq
where ${\hat \psi}_{\sigma}({\bf r})$ is the fermionic 
field operator of spin $\sigma$ and 
$g = 4\pi a/(1-(2ak_{cut}/\pi))$ is the bare coupling constant of the 
Fermi pseudo-potential interaction, with $k_{cut}$ the 
ultraviolet wavenumber cutoff \cite{braaten1,braaten2}. 
Braaten and Platter have also shown that the number 
${\cal N}_{pair}({\bf r})$ of pairs of fermions with opposite spins 
in a small ball of volume 
$4\pi s^3/3$ centered at the point ${\bf r}$ scales as 
$s^4{\cal C}({\bf r})/2$ for $s\to 0$, where 
${\cal C}({\bf r})=g^2 \langle {\hat \psi}_{\uparrow}^+({\bf r}) 
{\hat \psi}_{\downarrow}^+({\bf r}) 
{\hat \psi}_{\downarrow}({\bf r}) {\hat \psi}_{\uparrow}({\bf r}) 
\rangle$ is the contact density \cite{braaten1,braaten2}. 

Explicit expressions of the universal quantity $C$ have been 
derived by Tan \cite{tan1} 
for a uniform superfluid Fermi gas at zero temperature only in three limits: 
the Bardeen-Cooper-Schrieffer (BCS) limit of weakly bound Cooper pairs, 
the unitarity limit of infinite scattering length, 
and the Bose-Einstein condensate (BEC) limit of 
weakly-interacting bosonic molecular pairs. 

In this section we calculate the contact $C$ as a function of the 
inverse scattering parameter $1/(k_Fa)$ for a uniform superfluid 
Fermi gas in the full BCS-BEC crossover. 
To perform this calculation we use an efficient fitting 
formula of the ground-state 
energy \cite{nick,nick2} and the Tan's equation (\ref{figa}). 
We find that the contact $C$ has a maximum close 
to the unitarity limit of infinite scattering length. 
We also consider the interacting Fermi system 
under harmonic confinement. For this superfluid Fermi cloud we derive 
analytical formulas of the contact $C$ in 
the three relevant limits of the crossover. 

\subsection{Uniform superfluid Fermi gas at zero temperature} 

In the case of a zero-temperature uniform two-component 
superfluid Fermi gas of total density 
$n=n_{\uparrow}+n_{\downarrow}$ ($n_{\uparrow}=n_{\downarrow}$), 
large scattering length $a$ ($a\gg r_0$) in a volume $V$, 
the energy density can be written as 
\beq 
{E\over V} = {3\over 5} n \epsilon_F 
f({1\over k_F a}) \; , 
\label{miomao}
\eeq
where $f(y)$ is a universal function of inverse interaction 
parameter $y=1/(k_F a)$, with $\epsilon_F=\hbar^2k_F^2/(2m)$  
the Fermi energy and $k_F=(3\pi^2 n)^{1/3}$ 
the Fermi wave number\cite{leggett}. 
We observe that at finite finite temperature $T$ the function  
$f(y)$ is substituted by a more general universal function 
$\Phi(y,t)$, where $t=T/\epsilon_F$ is the scaled temperature. 
$\Phi(y,t)$ has been studied with Monte Carlo 
methods by Bulgac, Drut, and Migierski \cite{bulgac,bulgac01}, but only 
in the unitarity limit ($y=0$). 

It is straightforward to derive from Eq. (\ref{figa}) 
the expression of the contact density 
\beq 
{C\over V} = - {6\pi\over 5} k_F n {d f\over dy} \; .  
\label{cd}
\eeq
The behavior of $f(y)$ is well known in three relevant limits: 
\beq 
f(y) = \left\{ 
\begin{array}{cc}
1+ {10\over 9\pi} {1\over y} + O(1/y^2) \; , & y\ll -1    \\
\xi - \zeta y + O(y^2) \; , &  -1 \ll y \ll 1 \\
{5{\cal P}\over 18\pi}{1\over y} + O(1/y^{5/2}) \; , & y \gg 1
\end{array}
\right. 
\label{relevant}
\eeq 
In fact, in the weakly attractive limit ($y\ll -1$) 
one expects a BCS Fermi gas of weakly bound Cooper pairs.
As the superfluid gap correction 
is exponentially small, the function $f(y)$ follows 
the Fermi-gas expansion \cite{p15,p151}. 
In the so-called unitarity limit ($y=0$) one expects that 
the energy per particle is proportional to that of a non-interacting 
Fermi gas with the universal constant $\xi$ given by 
$\xi \simeq 0.42$ \cite{giorgini}. Note that more recent 
auxiliary-field Monte Carlo results \cite{gandolfi} 
predict a smaller value for $\xi$, namely $\xi \simeq 0.38$, 
while the experiment performed at Ecole Normale Superieure 
\cite{salomon} suggests $\xi \simeq 0.40$. The first correction 
to this behavior, shown in Eq. (\ref{relevant}), 
has been estimated from Monte Carlo data with $\zeta \simeq 1$ \cite{bertsch}. 
In the weak BEC limit ($y\gg 1$) 
one expects a weakly repulsive Bose gas of dimers. 
Such Bose-condensed molecules of mass $m_M=2m$ and 
density $n_M=n/2$  interact with a positive scattering length 
$a_M={\cal P} a$ with ${\cal P} \simeq 0.6$, 
as predicted by Petrov {\it et al.}\ \cite{p16}. 
In this regime, after subtraction of the molecular binding energy, 
the function $f(y)$ follows the Bose-gas expansion \cite{p17}.  
It is easy to obtain the contact intensity $C$ 
by using Eqs. (\ref{figa}) and (\ref{relevant}) 
in the relevant limits of the crossover. One finds 
\beq 
{C\over V} = \left\{ 
\begin{array}{cc}
{4\over 3} k_F^3 n a^2 + O(a^3) \; , & y\ll -1    \\
{6\pi\over 5} k_F n \zeta + O(1/a) \; , &  -1 \ll y \ll 1 \\
{{\cal P}\over 3} k_F^3 n a^2 + O(a^{7/2})  \; , & y \gg 1 
\end{array}
\right. 
\label{formulina}
\eeq 
in agreement with the previous determinations of Tan \cite{tan1} 
and Werner, Tarruell and Castin \cite{castin}. 
Notice that in the BEC limit we have removed the binding energy 
of molecules. Moreover, very recently finite temperature corrections 
to Eq. (\ref{formulina}) were given in Ref. \cite{finite-t}. 

\subsection{Contact intensity in the BCS-BEC crossover} 

Now we want to calculate the behavior of $C$ in the full BCS-BEC crossover. 
In 2005 we have proposed \cite{nick} the following 
analytical fitting formula 
\beq 
f(y) = \alpha_1 - \alpha_2 
\arctan{\left( \alpha_3 \; y \; 
{\beta_1 + |y| \over \beta_2 + |y|} \right)}  \; , 
\label{figata}
\eeq
interpolating the Monte Carlo energy per particle \cite{giorgini} 
and the limiting behaviors for large and small $|y|$. 
Eq. (\ref{figata}) is very reliable \cite{nick,nick2} and it 
has been successfully used by various authors 
for studying grund-state and collective properties of this 
superfluid Fermi system \cite{vari,vari2,vari3,vari4,vari5}. 
The parameter $\alpha_1$ is fixed by the value $\xi$ of $f(y)$
at $y=0$, the parameter $\alpha_2$ is fixed by the value of $f(y)$
at $y=\infty$, and $\alpha_3$ is fixed by the asymptotic $1/y$ coefficient
of $\epsilon(y)$ at large $|y|$.
The ratio $\beta_1/\beta_2$ is determined by the linear behavior $\zeta$
of $\epsilon(y)$ near $y=0$. 
The value of $\beta_1$ is then determined by minimizing the mean square 
deviation from the Monte-Carlo data. 
Of course, we have considered two different set of parameters: 
one set in the BCS
region ($y<0$) and a separate set in the BEC region ($y>0$) \cite{nick}. 
Table 1 of \cite{nick} reports the values of these parameters, 
with $\zeta=\zeta_{-}=1$ in the BCS region but $\zeta=\zeta_{+}=1/3$ in 
the BEC region. 

\begin{figure}
\begin{center}
\includegraphics[height=3.in,clip]{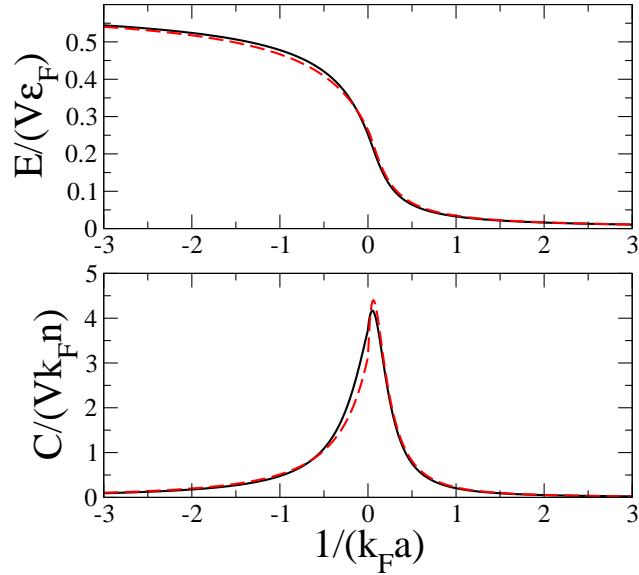}
\end{center}
\caption{Upper panel: Scaled ground-state energy $E/(V\epsilon_F)$ 
of the uniform Fermi gas as a function of the inverse interaction 
parameter $1/(k_Fa)$ in the BCS-BEC crossover. 
Lower panel: Scaled contact intensity $C/(Vk_F n)$ of the Fermi system 
as a function of the inverse interaction parameter $1/(k_Fa)$. 
Here $V$ and $n=N/V$ are the volume and the density 
of the Fermi gas, $k_F=(3\pi^2 n)^{1/3}$ is the Fermi wave number 
and $\epsilon_F=\hbar^2k_F^2/(2m)$ is the Fermi energy. 
Two different parametrization of the universal function 
$f(y)$: solid lines are obtained with Eq. (\ref{figata}), while dashed 
lines are calculated using the Pade approximant of \cite{kim}.}
\label{fig.1}
\end{figure}

Here we use Eq. (\ref{figata}) to calculate the contact 
density given by Eq. (\ref{cd}), but contrary to \cite{nick}, we choose 
$\zeta_{+}=\zeta_{-}=\zeta=1$ to ensure the continuity of $f'(y)$ at $y=0$. 
Notice that the recent experimental results 
obtained at Ecole Normale Superieure \cite{salomon} 
indeed suggest the continuity of $f'(y)$ at $y=0$.
In this way, in the BEC region $\beta_2=0.1517$ while $\beta_1$ is 
unchanged (see Table. 1 of \cite{nick}). 
In the upper panel of Fig. 1 we plot the ground-state energy $E$ 
while in the lower panel we plot the contact $C$, both 
as a function of the inverse interaction parameter $y=1/(k_Fa)$. 
For comparison, in addition to the data obtained 
with our method (solid lines), we insert also the results (dashed lines) 
one obtains using the Pade parametrization of $f(y)$ proposed 
by Kim and Zubarev \cite{kim}. The figure shows that 
solid and dashed lines are always close each other, 
apart for $-1\ll y \le 0$ where our fitting formula is 
smoother (and closer to the Monte Carlo data \cite{giorgini}). 
Moreover, the scaled contact $C/(Vk_Fn)$ as a function of $1/(k_Fa)$ 
has its maximum near to the unitarity limit $1/(k_Fa)=0$: 
the position of the maximum is located at 
$1/(k_Fa) \simeq 0.05$. Remarkably, the scaled contact 
has a behavior quite similar to the Landau's critical velocity $v_c$ 
(at which there is the breaking of superfluid motion), 
calculated along the BCS-BEC crossover. In fact, 
also $v_c$ goes to zero for $y\to \pm \infty$ and it 
has a peak at $y\simeq 0.08$ \cite{comb}. 
Clearly, the contact $C$ exhibits a maximum close to unitarity 
because we have subtracted the molecular binding energy contribution, 
given by $y^2$ in the BEC side ($y>0$). 
Including this energy term it is easy to show 
that $C$ increases monotonically from the BCS side to the BEC side, 
according the definition in Eq. (\ref{figa}). 
Nevertheless, as prevously stated, the radio-frequency spectroscopy 
shift is proportional to $C$, and its maximum around unitarity 
has been shown in Ref. \cite{zwierlein}, where the same artificial 
subtraction has been adopted. 

\subsection{Trapped superfluid Fermi gas} 

Let us now consider the superfluid Fermi gas 
under an external harmonic confinement 
\beq 
U({\bf r}) = {1\over 2} m \omega^2 (x^2 + y^2 + z^2) \; . 
\label{harmonic}
\eeq
In the limit of a large number $N$ of fermions we can use the 
local density (Thomas-Fermi) approximation 
\cite{flavio,flavio2,flavio3,flavio4,flavio5,flavio6,flavio7,flavio8} 
and the energy of the system can be written as 
\beq 
E = \int \left\{ {3\over 5}  n({\bf r}) \epsilon_F({\bf r}) 
f({1\over k_F({\bf r}) a}) + n({\bf r}) U({\bf r}) \right\} \ d^3{\bf r} \ , 
\label{functional}
\eeq
where $\epsilon_F({\bf r})=\hbar^2k_F({\bf r})^2/(2m)$ is the local  
Fermi energy and $k_F=(3\pi^2 n({\bf r}))^{1/3}$ 
is the local Fermi wave number. As in the 
uniform case, this expression is not very useful without 
the knowledge of the universal function $f(y)$. 

The numerical calculation of the contact indensity $C$ 
for a harmonically trapped Fermi superfluid 
in the full BCS-BEC crossover 
by using Eqs. (\ref{figa}) and (\ref{figata}) is very demanding. 
Consequently, we calculate $C$ only in the three relevant 
limits of the BEC-BEC crossover. In these limits we obtain useful 
analytical expressions for the contact $C$. 

\noindent 
{\bf BCS limit}. In the BCS limit ($a\to 0^{-}$) from Eqs. (\ref{relevant}) and 
(\ref{functional}) we find 
\beq 
{d E\over da} = {3\over 5} {10\over 9\pi} \int \left\{ 
\epsilon_F({\bf r}) n({\bf r}) k_F({\bf r}) \right\} \ d^3{\bf r} \; , 
\label{nice}
\eeq
where $n({\bf r})$ is the density profile of the ideal Fermi gas 
in the harmonic potential (\ref{harmonic}), given by \cite{ideal}
\beq 
n({\bf r}) = {2\sqrt{2}\over 3\pi^2 a_H^3} (3N)^{1/2} 
( 1 - {r^2\over r_F^2})^{3/2} \; , 
\eeq
where $r_F=\sqrt{2}(3N)^{1/6}a_H$ is the Fermi radius of the cloud, 
with $a_H=\sqrt{\hbar/(m\omega)}$ the characteristic harmonic length. 
Inserting this density profile into Eq. (\ref{nice}) 
and using Eq. (\ref{figata}) we get the contact intensity 
in the BCS limit: 
\beq 
C = {4096 \sqrt{2} \over 2835 \pi} {1\over a_H} 
\left({a\over a_H}\right)^2 (3N)^{3/2} \; . 
\eeq

\noindent 
{\bf Unitarity limit}. 
In the unitarity limit ($a\to \pm \infty $) from Eqs. (\ref{relevant}) and 
(\ref{functional}) we have 
\beq 
{d E\over da} = {3\over 5} {\zeta\over a^2} \int \left\{ 
\epsilon_F({\bf r}) n({\bf r}) {1\over k_F({\bf r})} \right\} 
\ d^3{\bf r} \; , 
\label{fine}
\eeq
where $n({\bf r})$ is the density profile of the unitary Fermi gas 
in the potential (\ref{harmonic}), namely \cite{flavio}
\beq 
n({\bf r}) =  {2\sqrt{2}\over 3\pi^2 a_H^3 \xi^{3/4}} (3N)^{1/2} 
( 1 - {r^2\over r_F^2})^{3/2} \; , 
\eeq
where $r_F=\sqrt{2}\xi^{1/4} (3N)^{1/6}a_H$ is the Fermi radius of the 
unitary cloud. Inserting this density profile into Eq. (\ref{fine})
and using Eq. (\ref{figata}) we obtain the contact in the unitarity limit: 
\beq 
C = {512 \sqrt{2}\over 525} {\zeta\over \xi^{1/4}} 
{1\over a_H} (3N)^{7/6} \; . 
\eeq

\noindent
{\bf BEC limit}. 
In the BEC limit ($a\to 0^+$) it is straightforward to calculate 
the contact $C$. In fact, the explicit formula of the ground-state energy 
of the dilute BEC is well known \cite{review-bose} and 
for a BEC of molecules it is given by 
\beq 
E = {5\over 7} {\hbar \omega\over 2} 
\left({15 {\cal P} a\over a_H} \right)^{2/5} 
\left({N\over 2}\right)^{7/5} \; ,   
\eeq 
where ${\cal P}\simeq 0.6$ and $N/2$ is the number of molecules. 
Then from Eq. (\ref{figa}) the contact intensity reads  
\beq 
C = {2\pi \over 7}{1\over a_H} 
\left({15 {\cal P}\over 2}\right)^{2/5} 
\left({a N\over a_H}\right)^{7/5} \; . 
\eeq

\section{Extended superfluid hydrodynamics}

In this section we discuss the extended Lagrangian density 
of superfluids we have proposed few years ago \cite{flavio} 
and applied to study mainly the unitary Fermi gas 
\cite{flavio,flavio2,flavio3,flavio4,flavio5,flavio6,flavio7,flavio8}. 
The internal energy density of this Lagrangian 
contains a term proportional to the kinetic
energy of a uniform non interacting gas of fermions, plus 
a gradient correction of the von-Weizsacker 
form $\lambda \hbar^2/(8m) (\nabla n/n)^2$ \cite{von}. 
This approach has been adopted for studying the quantum hydrodynamics 
of electrons by
March and Tosi \cite{tosi}, and by Zaremba and Tso \cite{tso}.
In the context of the BCS-BEC crossover, the gradient term 
is quite standard \cite{nick,kim,v2,v3,v4,v5,v6,v7,v8}. 
In particular we show the relation between our approach and 
the effective theory for the Goldstone field 
derived by Son and Wingate \cite{son}, and improved by 
Manes and Valle \cite{valle}, 
on the basis of conformal invariance. Finally, by using our equations 
of extended superfluid hydrodymics at zero temperature 
we calculate sound waves, static response function and 
structure factor of a generic superfluid.

The extended Lagrangian density of superfluids is given by 
\beq 
{\cal L} = - \hbar \, {\dot \theta} \, n - 
{\hbar^2\over 2m} ({\boldsymbol \nabla} \theta)^2 \, n 
- {\cal E}(n,{\boldsymbol\nabla} n) - U({\bf r}) \, n \; ,  
\label{lagrangian}
\eeq
where $n({\bf r},t)$ is the local density, $U({\bf r})$ is the 
external potential acting on particles, and $m$ the mass of 
superfluid particles. In the case of superfluid bosons 
$\theta({\bf r},t)$ is the phase of the condensate order parameter,  
while in the case of superfluid fermions 
$\theta({\bf r},t)$ is half of the phase of the 
condensate order parameter (of Cooper pairs). 
${\cal E}(n,{\boldsymbol\nabla} n)$ is the internal energy density 
of the system. Note that we are supposing that this equation of state 
${\cal E}(n,{\boldsymbol\nabla} n)$ can depend not only 
on the local density  $n({\bf r},t)$ but also on its space derivatives. 
For this reason we call (\ref{lagrangian}) the extended superfluid Lagrangian. 
We stress that in the context of the BCS-BEC crossover 
the extended internal energy density could be written as 
\beq 
{\cal E}(n,{\boldsymbol\nabla} n) = {\cal E}_0(n) + 
\lambda {\hbar^2\over 8m} {({\boldsymbol\nabla} n)^2\over n} \; , 
\eeq
where 
\beq 
{\cal E}_0(n) = {3\over 5} n \epsilon_F 
f({1\over k_F a})
\label{bonato}
\eeq
is the energy density discussed in the previous section 
(see Eq. (\ref{miomao})) which depends on 
the universal function $f(y)$ of the BCS-BEC crossover, 
while the second term is the gradient correction of the von-Weizsacker form 
\cite{flavio2}. In the BCS-BEC crossover we expect that 
$1/6 \leq \lambda \leq 1/4$, where $\lambda = 1/6$ 
is the appropriate value in BCS regime of weakly-interacting 
superfluid fermions of mass $m$ \cite{tosi,tso}, 
while $\lambda =1/4$ is the appropriate value 
in the deep BEC regime of  weakly-interacting 
superfluid bosonic dimers of mass $2m$ \cite{flavio2}. 

By using the Lagrangian density (\ref{lagrangian})  
the Euler-Lagrange equation 
\beq 
{\partial {\cal L} \over \partial \theta} - {\partial \over \partial t} 
{\partial {\cal L} \over \partial {\dot \theta}} - {\boldsymbol\nabla}\cdot 
{\partial {\cal L} \over \partial ({\boldsymbol\nabla} \theta) } = 0 
\eeq
gives 
\beq 
{\partial n \over \partial t} + {\hbar\over m} {\boldsymbol\nabla} \cdot 
\left( n \; {\boldsymbol\nabla} \theta \right) = 0 \; . 
\label{el1}
\eeq
The Euler-Lagrange equation 
\beq 
{\partial {\cal L} \over \partial n} - {\partial \over \partial t} 
{\partial {\cal L} \over \partial {\dot n}} - {\boldsymbol\nabla}\cdot 
{\partial {\cal L} \over \partial ({\boldsymbol\nabla} n)} = 0 
\eeq
gives instead 
\beq 
\hbar \, {\dot \theta} + {\hbar^2\over 2m}({\boldsymbol\nabla}\theta)^2 
+ U({\bf r}) + X(n,{\boldsymbol \nabla}n) = 0 \; ,  
\label{careful}
\eeq
where 
\beq 
X(n,{\boldsymbol \nabla}n) = 
{\partial {\cal E} \over \partial n} -  
{\boldsymbol\nabla}\cdot
{\partial {\cal E} \over \partial ({\boldsymbol\nabla} n)} \; 
\label{x-def}
\eeq
is the local chemical potential of the system 
(see also \cite{son} and \cite{valle}). 

The local field velocity ${\bf v}({\bf r},t)$ of the superfluid is 
related to the phase $\theta({\bf r},t)$ of the condensate by 
\beq 
{\bf v}({\bf r},t) = {\hbar \over m} 
{\boldsymbol \nabla} \theta({\bf r},t) \; . 
\label{velocity}
\eeq
This definition ensures that the velocity is irrotational, i.e. 
${\boldsymbol\nabla} \wedge {\bf v} = {\bf 0}$. 
By using the definition (\ref{velocity}) in both Eqs. (\ref{el1}) 
and (\ref{careful}) 
and applying the gradient operator $\boldsymbol\nabla$ 
to Eq. (\ref{careful}) one finds the extended hydrodynamic 
equations of superfluids
\beqa 
{\partial n \over \partial t} + {\boldsymbol\nabla} \cdot 
\left( n \; {\bf v} \right) = 0 \; . 
\\
m {\partial {\bf v} \over \partial t} + 
{\boldsymbol\nabla} \left[ {1\over 2} m {\bf v}^2 + U({\bf r}) + 
X(n,{\boldsymbol \nabla} n) \right] = {\bf 0} \; . 
\label{motion}
\eeqa

We stress that in the presence of an external confinement $U({\bf r})$ 
the chemical potential $\mu$ of the system does not coincide 
with the local chemical potential $X(n,{\boldsymbol \nabla} n)$. 
The chemical potential $\mu$ can be obtained from Eq. (\ref{careful}) 
setting $\theta({\bf r},t)=-\mu t/\hbar$ 
and ${\bf v}({\bf r},t)={\bf 0}$, such that 
\beq  
U({\bf r}) + X(n_0,{\boldsymbol \nabla}n_0) = \mu \; ,  
\eeq
where $n_0({\bf r})$ is the ground-state local density. 

The Lagrangian density (\ref{lagrangian}) depends on the dynamical variables 
$\theta({\bf r},t)$ and $n({\bf r},t)$. 
The conjugate momenta of these dynamical variables are then given by 
\beqa 
\pi_{\theta} &=& {\partial {\cal L} \over \partial {\dot \theta}} 
= - \hbar \, n \; , 
\\
\pi_{n} &=& {\partial {\cal L} \over \partial {\dot n}} = 0 \; , 
\eeqa
and the corresponding Hamiltonian density reads 
\beq 
{\cal H} = \pi_{\theta} \,  {\dot \theta} + \pi_{n} \,  {\dot n} 
- {\cal L} =  - \hbar \, n \,  {\dot \theta} - {\cal L} \; , 
\eeq
namely 
\beq 
{\cal H} = {\hbar^2\over 2m} ({\boldsymbol \nabla} \theta)^2 \, n 
+ {\cal E}(n,{\boldsymbol\nabla} n) + U({\bf r}) \, n \; ,  
\eeq
which is the sum of the flow kinetic energy density 
$\hbar^2({\boldsymbol \nabla} \theta)^2n/(2m)
= (1/2)m v^2n$, the internal energy density 
$ {\cal E}(n,{\boldsymbol\nabla} n)$, and the external energy 
density $ U({\bf r})n$.  

\subsection{Extended hydrodynamics in terms of Goldstone field}

Note that taking into account Eq. (\ref{careful}) one immediately finds 
\beq 
X(n,{\boldsymbol \nabla}n) \, n 
= - \hbar \, {\dot \theta} \, n - 
{\hbar^2\over 2m} ({\boldsymbol \nabla} \theta)^2\, n -  
U({\bf r})\, n \; . 
\eeq
Consequently the Lagrangian density (\ref{lagrangian}) can be rewritten as 
\beq 
{\cal L} = X(n,{\boldsymbol \nabla}n)\, n -  
{\cal E}(n,{\boldsymbol\nabla} n) \; .   
\eeq
Remarkably 
\beq 
P(n,\nabla n) = 
X(n,{\boldsymbol \nabla}n)\, n - {\cal E}(n,{\boldsymbol\nabla} n) 
\label{pressure}
\eeq
is the local pressure of the system 
as a function of the density and its spatial derivatives, 
which can be written as a function of the local chemical 
potential $X$ and its spatial derivatives, namely 
\beq 
{\cal L} = P(X,{\boldsymbol \nabla}X) \;  . 
\label{l-gold}
\eeq
This result, based on a Legendre transformation, 
is clearly illustrated in the book of Popov \cite{popov} 
and used in the recent papers of Son and Wingate \cite{son} and 
Manes and Valle \cite{valle}. 
Finally one can introduce the Goldstone field $\phi({\bf r},t)$ as 
\beq 
\phi({\bf r},t) = \theta({\bf r},t) + {\mu\over \hbar} t  \; . 
\eeq
In this way, by using again Eq. (\ref{careful}), one can write 
\beq 
X =  \mu  - \hbar {\dot \phi} - 
{\hbar^2\over 2m} ({\boldsymbol \nabla} \phi)^2 - 
U({\bf r})\; ,  
\eeq
Thus, the Lagrangian density (\ref{l-gold}) actually depends only on the 
Goldstone field $\phi({\bf r},t)$. This is exactly the 
main message of the paper of Son and Wingate \cite{son}, which 
however traces back to the older results of Popov \cite{popov}. 

\subsection{Application: the Unitary Fermi gas}

Let us now suppose that the equation of state of the superfluid 
at zero temperature is that of the unitary Fermi gas, i.e. 
\beq 
{\cal E}(n,{\boldsymbol\nabla} n) = {3\over 5} {\hbar^2\over 2m} 
(3\pi^2)^{2/3} \xi \, n^{5/3} + 
\lambda {\hbar^2\over 8m} {({\boldsymbol\nabla} n)^2\over n} \; ,  
\eeq
where $\xi \simeq 0.4$ and $\lambda \simeq 0.25$ \cite{flavio4,flavio8}. 
It follows from Eq. (\ref{x-def}) that 
\beq 
X(n,{\boldsymbol\nabla} n) =  {\hbar^2\over 2m} 
(3\pi^2)^{2/3} \xi \, n^{2/3} -  
\lambda {\hbar^2\over 8m} {({\boldsymbol\nabla} n)^2\over n^2} \; , 
\label{invert}
\eeq
by taking into account that the surface terms give zero contribution. 
In addition we get from Eq. (\ref{pressure}) that 
\beq 
P(n,{\boldsymbol\nabla} n) =  {2\over 5}{\hbar^2\over 2m} 
(3\pi^2)^{2/3} \xi \, n^{5/3} -  
\lambda {\hbar^2\over 4m} {({\boldsymbol\nabla} n)^2\over n} \; , 
\eeq
The Lagrangian density of Eq. (\ref{l-gold}) is then obtained 
by finding $n$ and $\nabla n$ as functions of $X$ and $\nabla X$ 
by inverting Eq. (\ref{invert}). 
This can be done in terms of a derivative expansion. 
One gets $n = (2m \xi)^{3/2} 
X^{3/2}/(3\pi^2\hbar^3)$ and 
\beq 
{\cal L} = {\cal L}_{LO}+ {\cal L}_{NLO} \; 
\label{lmio}
\eeq
where 
\beq 
{\cal L}_{LO} = c_0 \, {m^{3/2}\over \hbar^3} \, X^{5/2} \; ,  
\eeq
with 
\beq 
c_0 = {2^{5/2}\over 15\pi^2 \xi^{3/2}} \; , 
\eeq
is the Lagrangian density at the leading order, and 
\beq 
{\cal L}_{NLO} = c_1 \, {m^{1/2}\over \hbar} \, 
{(\nabla X)^2\over \sqrt{X}} \; ,  
\eeq
with 
\beq
c_1 = -\lambda {3 \cdot 2^{1/2}\over 8\pi^2 \xi^{3/2}} \; ,  
\eeq
is the next-to-leading contribution to the Lagrangian density. 
The Lagrangian density (\ref{lmio}) is the same of that 
derived by Son and Wingate \cite{son} from general coordinate 
invariance and conformal invariance. Actually, at the lext-to-leading 
order Son and Wingate have found an additional term \cite{son}, 
which has been questioned by Manes and Valle \cite{valle} and 
is absent in our approach. 

\subsection{Nonlinear sound waves, static response function 
and structure factor} 

In this subsection we consider the following zero-temperature 
equation of state of a generic superfluid 
\beq 
{\cal E}(n,{\boldsymbol\nabla} n) = {\cal E}_0(n) + 
\lambda {\hbar^2\over 2m} {({\boldsymbol\nabla} n)^2\over 4n} \; .   
\eeq
Here the internal energy is the sum of two contributions: 
a generic internal energy ${\cal E}_0(n)$ which depends only 
of the local density $n({\bf r},t)$ (for instance that of Eq. (\ref{bonato})) 
plus the gradient correction of the von Weizsa\"cker type, where 
the coefficient $\lambda$ can be a function of the interaction strength. 

The equation of motion (\ref{motion}) becomes 
\beq 
m {\partial {\bf v} \over \partial t} + 
{\boldsymbol\nabla} \left[ {1\over 2} m {\bf v}^2 + U({\bf r}) + 
X_0(n) - \lambda {\hbar^2\over 2m} 
{\nabla^2\sqrt{n}\over \sqrt{n}}
\right] = {\bf 0} \; , 
\eeq
with 
\beq 
X_0(n) = {\partial {\cal E}_0 \over \partial n} \; . 
\eeq

We are interested on the propagation of sound waves in superfluids. 
For simplicity we set 
\beq 
U({\bf r})= 0 \; , 
\eeq 
and consider a small perturbation ${\tilde n}({\bf r},t)$ around 
a uniform and constant configuration $n_0$, namely 
\beq 
n({\bf r},t) = n_0 + {\tilde n}({\bf r},t) \; . 
\eeq
Neglecting quadratic terms in ${\tilde n}$ and $v$ we derive the 
linearized equations of extended superfluid hydrodynamics 
\beqa 
{\partial {\tilde n} \over \partial t} 
+ n_0 {\boldsymbol\nabla} \cdot {\bf v} = 0 \; ,  
\label{pp1}
\\
n_0 {\partial {\bf v} \over \partial t} + 
{c_s^2} {\boldsymbol\nabla} {\tilde n}
- \lambda {\hbar^2\over 4m^2} {\boldsymbol\nabla} 
\left(\nabla^2 {\tilde n}\right)  = {\bf 0} \; , 
\label{pp2}
\eeqa
where $c_s$ is the sound velocity, given by 
\beq 
c_s^2 = {n_0\over m} {\partial {X}_0(n_0)
\over \partial n} = 
{n_0\over m} {\partial^2 {\cal E}_0(n_0)\over \partial n^2} \; .  
\eeq
Applying the operator ${\partial\over\partial t}$  
to Eq. (\ref{pp1}) and the operator ${\boldsymbol\nabla}$ 
to Eq. (\ref{pp2}) and subtracting the two resulting equations we get 
\beq 
\left( {\partial^2\over\partial t^2} - c_s^2 \nabla^2 
+ \lambda {\hbar^2\over 4m^2}  \nabla^4 \right) 
{\tilde n}({\bf r},t) = 0 \; . 
\eeq
This is the wave equation of the small perturbation 
${\tilde n}({\bf r},t)$ around the uniform density $n_0$. 
We stress that the effect of the gradient term in the equation of state 
is the presence of quartic spatial derivatives in this wave equation. 

It is straightforward to show that the wave equation admits 
the real solution 
\beq 
{\tilde n}({\bf r},t) = A \, e^{i({\bf q}\cdot {\bf r}-\omega_q t)}  
+ A^* \, e^{-i({\bf q}\cdot {\bf r}-\omega_q t)} \; , 
\eeq
where the frequency $\omega_q$ and the wave vector ${\bf q}$ are 
related by the Bogoliubov-like dispersion formula 
\beq 
\hbar \omega_q = \sqrt{{\hbar^2q^2\over 2m} 
\left( \lambda {\hbar^2q^2\over 2m} + 2 m c_s^2 \right) } \; ,    
\eeq
or equivalently 
\beq 
\omega_q = c_s q \sqrt{1+\alpha q^2} = c_s q 
\left( 1 + {\alpha \over 2}q^2 - {\alpha^2\over 8} q^4 + ... 
\right) \; , 
\eeq
with $\alpha=\lambda \hbar^2/(4m^2c_s^2)$. Thus, the dispersion relation 
$\omega_q$ is linear in $q=|{\bf q}|$ only for small values of the 
wavenumber $q$ and becomes quadratic for large values of $q$. 

We observe that, for a generic many-body system, 
the dispersion relation can be written as \cite{treiner}
\beq 
\hbar \omega_q = \sqrt{m_1(q)\over m_{-1}(q)} 
\label{e59}
\; , 
\eeq
where $m_n(q)$ is the $n$ moment of the dynamic structure function 
$S(q,\omega)$ of the many-body system under investigation, i.e. 
\beq 
m_n(q) = \int_0^{\infty} d\omega \ S(q,\omega) \ (\hbar \omega)^n \; . 
\eeq 
Note that Eq. (\ref{e59}) is not exact and is valid under the approximation 
of a single-mode density excitation. It therefore only gives an upper bound 
for the dispersion relation. This is important for large $q$, for
which quasiparticles other than density excitations may contribute. 
In our problem we have 
\beq 
m_1(q) = {\hbar^2q^2\over 2m}
\eeq
and 
\beq 
m_{-1}(q) = {1 \over  \lambda {\hbar^2q^2\over 2m} + 2 m c_s^2} \; .  
\eeq

In general, the static response function $\chi(q)$ is defined as \cite{treiner}
\beq 
\chi(q) = -2 \ m_{-1}(q) \; , 
\eeq 
and in our problem it reads 
\beq 
\chi(q) = - {2 \over  \lambda {\hbar^2q^2\over 2m} + 2 m c_s^2} \; ,   
\eeq
or equivalently 
\beq 
\chi(q) = - {1 \over  m c_s^2} {1\over 1+\alpha q^2} = 
- {1 \over  m c_s^2} \left( 1 - \alpha 
q^2 + \alpha^2 q^4 + ... \right) \; ,   
\eeq
where again $\alpha=\lambda \hbar^2/(4m^2c_s^2)$. 

The static structure factor $S(q)$ is instead defined as \cite{treiner}
\beq 
S(q) = m_0(q) = \int_0^{\infty} d\omega \ S(q,\omega) \; ,   
\eeq
but it can be approximated by the expression 
\beq 
{\tilde S}(q) = \sqrt{m_1(q)\, m_{-1}(q)} \; , 
\eeq
which gives an upper bound of $S(q)$, i.e. ${\tilde S}(q)\geq S(q)$ 
\cite{treiner}. 
In our problem we immediately find 
\beq 
{\tilde S}(q) =  \sqrt{{\hbar^2q^2\over 2m} 
\over  \lambda {\hbar^2q^2\over 2m} + 2 m c_s^2} \; ,  
\eeq
or equivalently 
\beq 
{\tilde S}(q) = {\hbar q \over 2 mc_s} {1\over \sqrt{1+\alpha q^2}} 
= {\hbar q \over 2 mc_s} \left( 1 - {1\over 2} \alpha q^2 + 
{3\over 8} \alpha^2 q^4 + ... \right) \; . 
\eeq
Our results clearly indicate that it should be possible to observe 
experimentally the effect of the dispersive von-Weizsacker-like gradient 
term from sound-wave measurements.  

\section{Conclusions}

In the first part of this contribution we have calculated 
the contact $C$ as a function of the 
inverse scattering parameter $1/(k_Fa)$ for a uniform superfluid 
Fermi gas in the full BCS-BEC crossover at zero temperature. 
We have found that the contact $C$ has a maximum close 
to the unitarity limit of infinite scattering length, in  
analogy with the behavior of the Landau's critical velocity $v_c$,  
at which there is the breaking of superfluid motion \cite{comb}. 
We have also considered the interacting Fermi system 
under harmonic confinement. In this case, we have derived 
analytical formulas of the contact intensity $C$ in 
the three relevant limits of the crossover. 
Our results can be experimentally tested with ultracold atomic clouds 
by measuring one of the quantities which are directly 
related to the contact intensity $C$: the tail of the momentum 
distribution, the derivative of the total energy with respect 
to the scattering length, the radio-frequency spectroscopy shift, or 
the photoassociation rate.  
In the second part we have analyzed some properties of 
the extended superfluid hydrodynamics \cite{flavio}; in particular, 
we have shown its strict relation with the low-energy effective field theory 
built on the Goldstone mode. Finally, by using 
the extended hydrodynamics we have calculated, for generic superfluid 
in the absence of external confinement, the nonlinear dispersion relation 
of sound waves, and, as a by-product, both static response function and 
structure factor.

\end{document}